\documentclass[10pt,journal,compsoc]{IEEEtran}
\ifCLASSOPTIONcompsoc
  \usepackage[nocompress]{cite}
\else
  \usepackage{cite}
\fi
\ifCLASSINFOpdf
   \usepackage[pdftex]{graphicx}
\else
\fi
\usepackage{array}
\usepackage{url}

\newcommand{\insertpic}[1]{\includegraphics[width=0.9\columnwidth]{#1}}
\newcommand{\etal}{\textit{et al.}}
\newcommand{\funcname}[1]{\texttt{#1}}

\begin{document}
%
\title{Software Reliability Growth Models Predict Autonomous Vehicle Disengagement Events}
%
%
%
%

\author{Robert~Merkel
\IEEEcompsocitemizethanks{\IEEEcompsocthanksitem R. Merkel is with the Faculty
of Information Technology, Monash University, Melbourne, Victoria, Australia.\protect\\
E-mail: robert.merkel@monash.edu}}

%
%

\markboth{}
{Merkel: Software Reliability Growth Models...Disengagement Events}
%



\IEEEtitleabstractindextext{%
\begin{abstract}
The acceptance of autonomous vehicles is dependent on the rigorous
assessment of their safety.  Furthermore, the commercial viability
of AV programs depends on the ability to estimate the time and resources required
to achieve desired safety levels.  Naive approaches to estimating the reliability and safety levels of autonomous vehicles under development are will require infeasible amounts of testing
of a static vehicle configuration.  To permit both the estimation of current safety, and make predictions about the reliability of future systems, I propose the use of a standard tool for modelling the reliability of evolving software systems, software reliability growth models (SRGMs).

Publicly available
data from Californian public-road testing of two autonomous vehicle systems is
modelled using two of the best-known SRGMs.  The ability of the models to accurately estimate current
relibility, as well as for current testing data to predict reliability in the future after
additional testing, is evaluated.
One of the models, the Musa-Okumoto model, appears to be a good estimator and a reasonable
predictor.
\end{abstract}

}

\maketitle

\IEEEdisplaynontitleabstractindextext

%
\IEEEpeerreviewmaketitle

\IEEEraisesectionheading{\section{Introduction}\label{sec:introduction}}

%
%
%
%

\IEEEPARstart{T}{he} safety and reliability of autonomous vehicles (AVs), which are currently
under development by a wide variety of companies, is of significant public importance.  Regulators in many jurisdictions are currently defining regulatory frameworks for the approval of autonomous vehicles to operate on public roads~\cite{noauthor_automated_nodate-1}.  Public acceptance of AV technology will depend on deployed AVs being safer for both their passengers and other road users than current human-driven vehicles~\cite{liu_how_2018}.  Therefore regulatory and operator approval of AVs will require rigorous evidence that AV accident and particularly fatality rates are lower than acceptable thresholds.

Kalra and Paddock~\cite{kalra_driving_2016} have shown the infeasibility of a naive approach to demonstrating the safety of AV systems.  In short, if safety were demonstrated by testing a fixed AV system configuration in conditions reflecting
typical usage, a fleet of such vehicles would have to be driven 275 million miles (approximately 441 million kilometres) without a fatality for the probability that the fatality rate for that AV system was lower than for conventional vehicles to exceed 95\%.  Even ignoring the exorbitant cost of such a process, it is highly unlikely that the AV system software and hardware could truly be kept static for long enough to complete such a testing program.  Therefore, an alternative approach is required.

While external stakeholders are likely to be most interested in the reliability of a system
as it currently exists, AV manufacturers also have a considerable stake in predicting the reliability of their future AV systems \emph{before they are completed}.  AV development programs by major manufacturers have cost over one billion US dollars~\cite{harris_google_2017} and taken over a decade. The profitability of such investments depends on the time and resources required to build a commercially viable product.  Therefore, the ability to estimate future safety improvements would allow AV manufacturers to evaluate whether further investments are financially justifiable.

Evaluating and predicting  the reliability of an evolving software system is a well studied problem.  Software reliability growth models (SRGMs) have been developed for the purpose~\cite{wood_software_1996}.  SRGMs allow the statistically rigorous estimation of the current and future reliability of software systems as programming faults are rectified through testing and use.  Given that only a small minority of vehicle accidents are caused by hardware failures~\cite{rechnitzer_george_effect_2000}, it is plausible that failure rates of AVs can be modelled using techniques developed for software systems.

Previous analyses of accident data from on-road AV test programs have shown the counter-intuitive result that accident rates have not declined~\cite{favaro_examining_2017} over years of testing and development.  Aside from the very limited sample, this may be due to the fact
that a substantial proportion of accidents will be attributable to a varying degree to the actions of the drivers of other vehicles involved.  Attributing responsibility in collisions between vehicles is self-evidently complex and may render any analysis less valid than might be hoped.

However, the publicly available reports of on-road testing by AV manufacturers to the California Department of Motor Vehicles~\cite{noauthor_testing_nodate} provide a very useful proxy metric for estimating the progress of an under-test AV system.  These reports, as well as listing accidents involving AVs under test, list each occasion where a \emph{disengagement events} occurred.  A disengagement event is defined as occuring when a human backup driver either:

\begin{itemize}
 \item takes over driving after a warning from the AV's systems that they were not able to proceed safely in accordance with local driving laws.
 \item takes over driving on their own initiative where they judged that the AV was not proceeding safely in accordance with local driving laws.
\end{itemize}

To achieve full autonomous operation (Levels 4 and 5 according to the widely-adopted SAE taxonomy~\cite{noauthor_taxonomy_2018}) the rate of these incidents will have
to be reduced to negligible levels -- even if the human backup drivers were ultimately overzealous in some interventions, customers are unlikely
to accept vehicles that put them in driving situations they would themselves consider too risky. It is also at least plausible that the rate of such events is reasonably well correlated with the rate at which accidents which the AV system could have prevented would occur in the absence of the backup driver.

Therefore, the disengagement rate is at least a plausible initial metric for assessing the reliability of AV systems.

In this paper, we examine:

\begin{itemize}
 \item if two well-known SRGMs accurately fit reported disengagement rate data for the two most extensively tested AV systems
 \item if the two SRGMs can be used to accurately predict disengagement rates by modelling using a subset of earlier data, and comparing the model predictions with the later data
 \item which of the two SRGMs is most useful for these purposes.
\end{itemize}

\section{Background}

\subsection{Software Reliability Modeling}

Software reliability is a measure of the frequency of \emph{failures} -- instances where a software system fails to perform as specified.  More formally, the IEEE Software Reliability recommended practice~\cite{noauthor_ieee_2017} defines software reliability as:

\begin{quote}
\begin{enumerate}
 \item The probability that software will not cause the failure of a system for a
specified time under specified conditions.
\item The ability of a program to perform a required function under
stated conditions for a stated period of time.
\end{enumerate}
\end{quote}

For our purposes, the first definition is the relevant one, though it is typical to describe reliability by measuring the presence of failures rather than their absence.

A software reliability model (SRM) is therefore a mathematical expression describing the reliability of a system, or more formally ``A mathematical expression that specifies the general form of the software failure process as a function of factors such as fault introduction, fault removal, and the operational environment''~\cite{noauthor_ieee_2017}.  In this context, faults are defined as the underlying defects in the software system that are the cause of failures.\footnote{The relationship between software faults and failures is a complex one discussed at length in the software engineering literature, but the
subtleties are not pertinent here.}

Software Reliability Growth Models (SRGMs) are a subset of SRMs, based on the observation of the distinct reliability characteristics of software-based systems compared to ones not including software.  In hardware reliability modelling, the
primary source of failure is physical deterioration, whereas in software the primary cause of failures is design faults, which, once fixed, do not recur~\cite[p. 7]{musa_john_d._software_1987}\footnote{to a first approximation; ``regression errors'' due to poor source code
change management are not uncommon, and bug fixes are often less than perfect}.  Therefore, software systems often demonstrate a characteristic pattern of reliability, where failures are common in early testing/use, but as the underlying errors
are fixed the rate of failure drops and reliability improves.

\begin{figure}[!h]
\insertpic{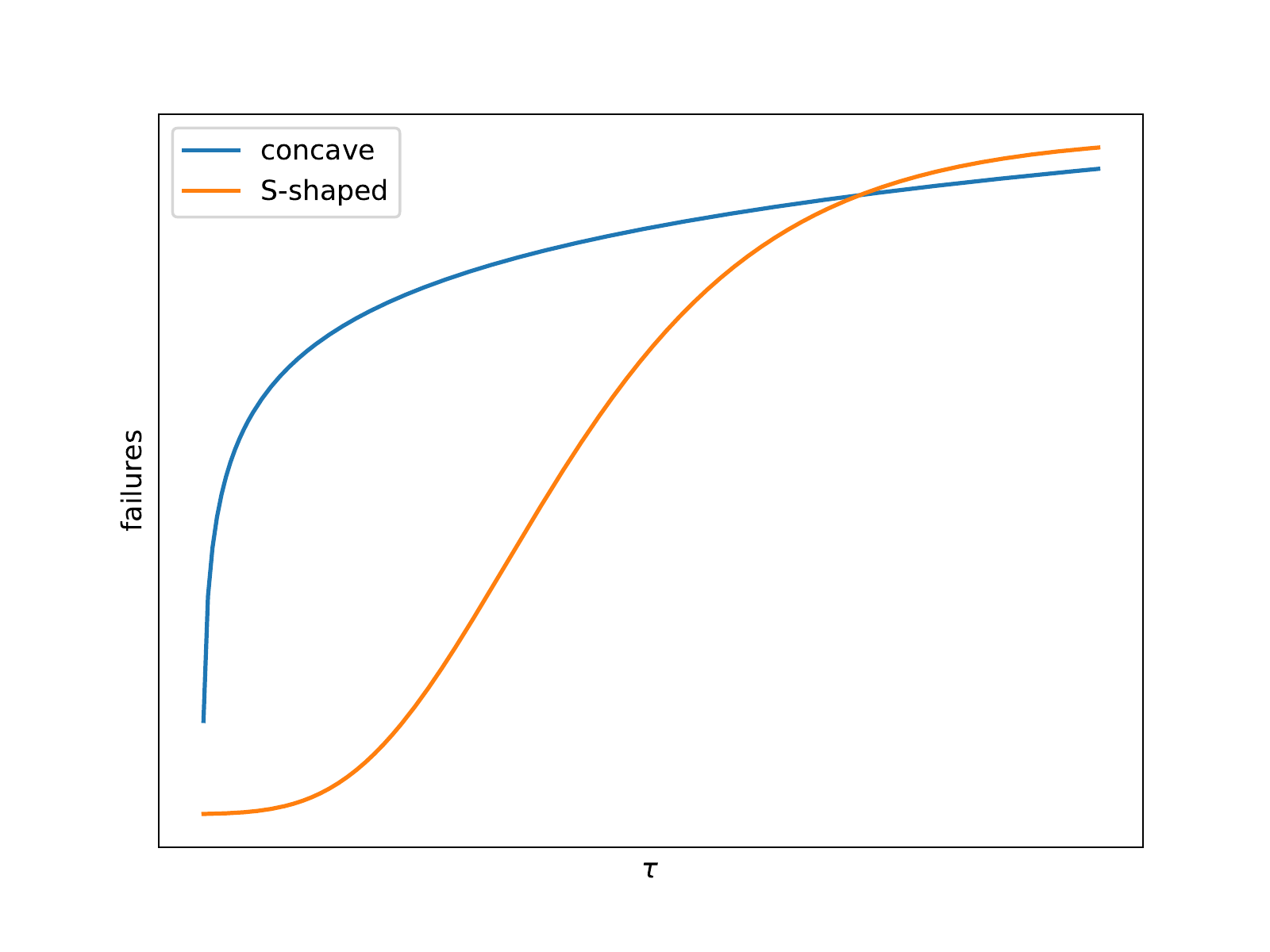}
\caption{{\bf Concave and S-shaped SRGMs.}
Representative examples of concave and s-shaped SRGMs, illustrating how S-shaped
SRGMs initially have a low rate of failure detection which increases then decreases.}
\label{fig0}
\end{figure}

A variety of SRGMs have been proposed, all making slightly different assumptions about the nature of software faults, the efficacy of bug fixing, and testing/usage patterns.  Most such models fall into one of two groups, S-shaped and concave~\cite{wood_software_1996}, based on the characteristic shape of the model when plotted.  Figure~\ref{fig0} shows an illustrative example of each model group.  The X-axis of the figure represents ``time'' $\tau$ (which may or may not be simple calendar time, as we will discuss further) and the Y-axis represents the cumulative number of failures detected from the commencement of testing until time $\tau$.  As can be seen, in the concave model, the rate of failure detection is highest at the beginning of data collection, and decreases as time goes on.  In S-shaped models, failure detection rates initially increase as the effectiveness of testing improves, before decreasing as more of the defects in the system are found.  In both cases, the rate of failure detection asymptotically approaches zero as $\tau$ approaches infinity.

Empirical studies of software failure data~\cite{wood_software_1996, ullah_comparative_2012} have shown that different model classes are better fits for different projects.  Ullah, Morisio, and Vetro~\cite{ullah_comparative_2012} studied the performance of a variety of such models using the histories of real software
development projects.  They found that the  Musa-Okumoto model (a concave model) and the Gompertz model (an S-shaped model) fitted empirical data more accurately than other models.  Therefore, for this preliminary study, those two models were selected.

\subsubsection{Musa-Okumoto model}

The Musa-Okumoto model~\cite{musa_logarithmic_1984} is a popular concave SRGM.  It is based on the following assumptions:

\begin{itemize}
 \item that failures occur as a nonhomogeneous Poisson process
 \item that the number of failures $\mu$ at the start of the process $\left(\tau = 0\right)$ is 0.
 \item that expected \emph{failure intensity} $\lambda$- the rate at which failures are expected to occur - declines exponentially with the number of failures detected ($\mu$).  More formally, $\lambda\left(\mu\right) = \lambda_{0} e^{-\theta \mu}$, where $\lambda_0$ is the failure intensity at the start of the process, and $\theta$ is a parameter that describes the rate of failure intensity decline.
\end{itemize}

In the Musa-Okumoto model, the \emph{mean value function} - the expected number of failures detected at time $\tau$ is designated $\mu\left(\tau\right)$, and is described by the following function:

\begin{equation}
 \mu\left(\tau\right) = \frac{1}{\theta} \ln\left(1 + \lambda_{0}\theta \tau\right)
\end{equation}

\subsubsection{Gompertz model}

The Gompertz model~\cite{huang_unified_2003} is an example of an S-shaped SRGM.  Gompertz growth models have been used to predict growth in a variety of domains, including microbiology as well as software engineering (need citation).  The mean value function of the Gompertz model is as follows:

\begin{equation}
 \mu \left(\tau \right) = a k^{b^{\tau}} , a > 0, 0 < b < 1, 0 < k < 1
\end{equation}

$a$ is the total number of failures to be ``eventually'' detected, and $b$ and $k$ are parameters to be estimated.

\section{Methods}

\subsection{Data selection}
This study uses the publicly available vehicle event reports from AV manufacturers available from the California Department of Motor Vehicles~\cite{noauthor_testing_nodate}.

These reports, which must be provided to the DMV on a yearly basis as a condition of the autonomous vehicle testing permits offered by that state, provide the provide the following information about each AV manufacturer's program:

\begin{enumerate}
 \item A listing of all the vehicles used in testing.
 \item the date, time, and brief description of the reasons for the disengagement.
 \item the total vehicle distance driven in autonomous mode per calendar month.
 \item total vehicle distance driven in autonomous mode per calendar month for each individual vehicle in the test program.
\end{enumerate}

Of the 20 AV manufacturers who submitted testing reports for the 2017 calendar year, only two were selected for analysis: Waymo~\cite{noauthor_waymo_nodate}, formerly Google's autonomous vehicle division, and Cruise Automation~\cite{noauthor_cruise_nodate}, the autonomous vehicle program of General Motors.  These two companies were selected because they have conducted far more on-road testing in California than the other 18 companies in the program: Waymo has completed over 1.4 million
miles of testing since commencing in 2014, and Cruise has completed approximately 141,000 miles of testing since commencing in 2015.  The next largest California-based testing program has completed less than 10,000 miles of on-road testing.

This selection is not intended as a judgement on the overall maturity and readiness of AV programs, as the other
AV manufacturers may be conducting large-scale testing on private roads or in other jurisdictions, only that there is insufficient data in the California data set to evaluate disengagement event trends for other manufacturers.

\subsection{Data preprocessing}

In this study, all disengagement events listed in the company reports were included in the analysis.  It is likely that in some cases, disengagements occurred as a precaution by the safety driver, and no adverse event would have resulted
if the AV continued in autonomous mode.  However, the information provided in the data set was insufficient to make such a judgement on whether to exclude any particular disengagement event on this basis.

When applying SRGMs such as Musa-Okumoto, ``time'' can be defined in a number of ways, including ``clock time'', CPU time spent executing the software, or the number of tests run.  In the context of AVs, the distance driven is the most natural measure of the testing conducted, particularly given that crash rates are typically quoted in terms of distance travelled.

Unfortunately, however, the exact cumulative distance driven at the time of each disengagement event is not reported in the public data set - only the number of miles driven by the driverless car fleet each calendar month.  The exact calendar time of each disengagement event is reported, but there is no information about the total fleet hours driven up to the time of each disengagement event.  Inspection of the full disengagement event logs and the accumulated testing
suggested that, at least in the early stages of development, on-road testing was irregular and conducted for only a small fraction of each calendar month.  This meant that, if calendar time was used as the time variable in modelling, one of the  assumptions
of the Musa-Okumoto model - that disengagements were a random memoryless Poisson process - did not hold in the data available.

Therefore, for each AV manufacturer, for each calendar month in the data set, we calculated:

\begin{itemize}
\item The cumulative miles driven by the manufacturer's vehicle fleet up to and including that calendar month; and
\item The cumulative number of disengagement events.
\end{itemize}

\subsection{Model parameter estimation}

Both the Musa-Okumoto and Gompertz models are parameterized, and to apply the model to a given software system values for these parameters must be estimated.  In a random process, any parameters chosen will usually not result in a perfect
fit, and therefore a criterion that specifies the nature of the ``best'' fit is required.  The two definitions of best fit commonly used for this type of estimation in the literature are maximum likelihood and nonlinear least squares.

Maximum likelihood estimation seeks to choose parameters that maximise the \emph{likelihood function} of the model.  The likelihood function $L\left(\theta\right)$ of a model with observed data $x$ is defined as ``the probability of the observed data $x$ considered as a function of $\theta$''~\cite{pawitan_yudi_all_2001}.  Therefore,
a maximum likelihood estimate chooses the model where the observed data was
most likely to have occurred.

Nonlinear least squares estimation seeks to find parameters that minimise the sum of the square of the differences between the produced data, and the values predicted by the model.

In both cases, iterative numerical methods are used to find parameters that minimise or maximise the relevant function.

In linear models with errors that are normally distributed, the best estimate obtained
by least squares will also be the best maximum likelihood estimate, but the model here
is not linear and there is no guarantee the errors are normally distributed.  The existing SRGM literature generally uses ML modelling where possible to obtain parameter
estimates.  Unfortunately, while the likelihood function of the Musa-Okumoto model is known, the closed forms in the literature require that the data must be in the form of intervals between individual events.

As the data was not available in this form, least squares estimation was used to estimate model parameters for both the Gompertz and Musa-Okumoto models.  The \funcname{scipy.optimize.curve\_fit} function in version 0.19 of SciPy~\cite{noauthor_scipy_2018}, using Python 3.6.7rc1 on an Ubuntu 18.10 virtual machine instance, was used for all curve fitting.

\funcname{curve\_fit} also estimates a covariance matrix, which provides variance estimates
for the model parameters.  The delta method~\cite{noauthor_how_nodate} was used to calculate 95\% confidence intervals for the model.

\subsection{Model accuracy assessment metrics}

Previous studies comparing the goodness of fit for various software reliability growth models~\cite{ullah_comparative_2012} have used the coefficient of determination ($R^{2}$) to compare the quality of the model fitting achieved.  While this metric
works very well for comparing linear models, it can give misleading results when comparing nonlinear models~\cite{spiess_evaluation_2010}.  The most sophisticated comparison metrics for nonlinear models require the ability
to calculate the likelihood functions for those models, which was not possible given the data available.

Therefore, for this study, the standard error of estimate~\cite{salkind_standard_2010}  was used to compare the goodness of fit of the models.  The standard error of estimate is the mean deviation of the observations from the model prediction.  The smaller the standard error of estimate, the better the model fits the data.

\subsection{Experimental procedure}

To compare the usefulness of the two models for AV data, we adapted the procedure of Ullah \etal~\cite{ullah_comparative_2012}.  As discussed, we use the standard error of estimate to compare the goodness of model fit, rather than coefficient of determination.

\subsubsection{Experiment 1: goodness of fit}

To assess which model best fitted the disengagement event data from each manufacturer, a nonlinear least squares fit was computed using the entire data set from each AV manufacturer, for both the Musa-Okumoto and the Gompertz models.  The standard error of estimate was computed for each model.

\subsubsection{Experiment 2: accuracy of predictions}

To assess which model most accurately predicts disengagement event rates, a nonlinear least squares fit was computed for the first two-thirds of the points of each data set from the AV manufacturer, for each model.  Both models were then plotted against the actual data for the entire data set.

The standard error of estimate, and 95\% model confidence intervals, were computed for each model.

\section{Results}

\subsection{Experiment 1: goodness of fit}

\begin{figure}[!h]
\insertpic{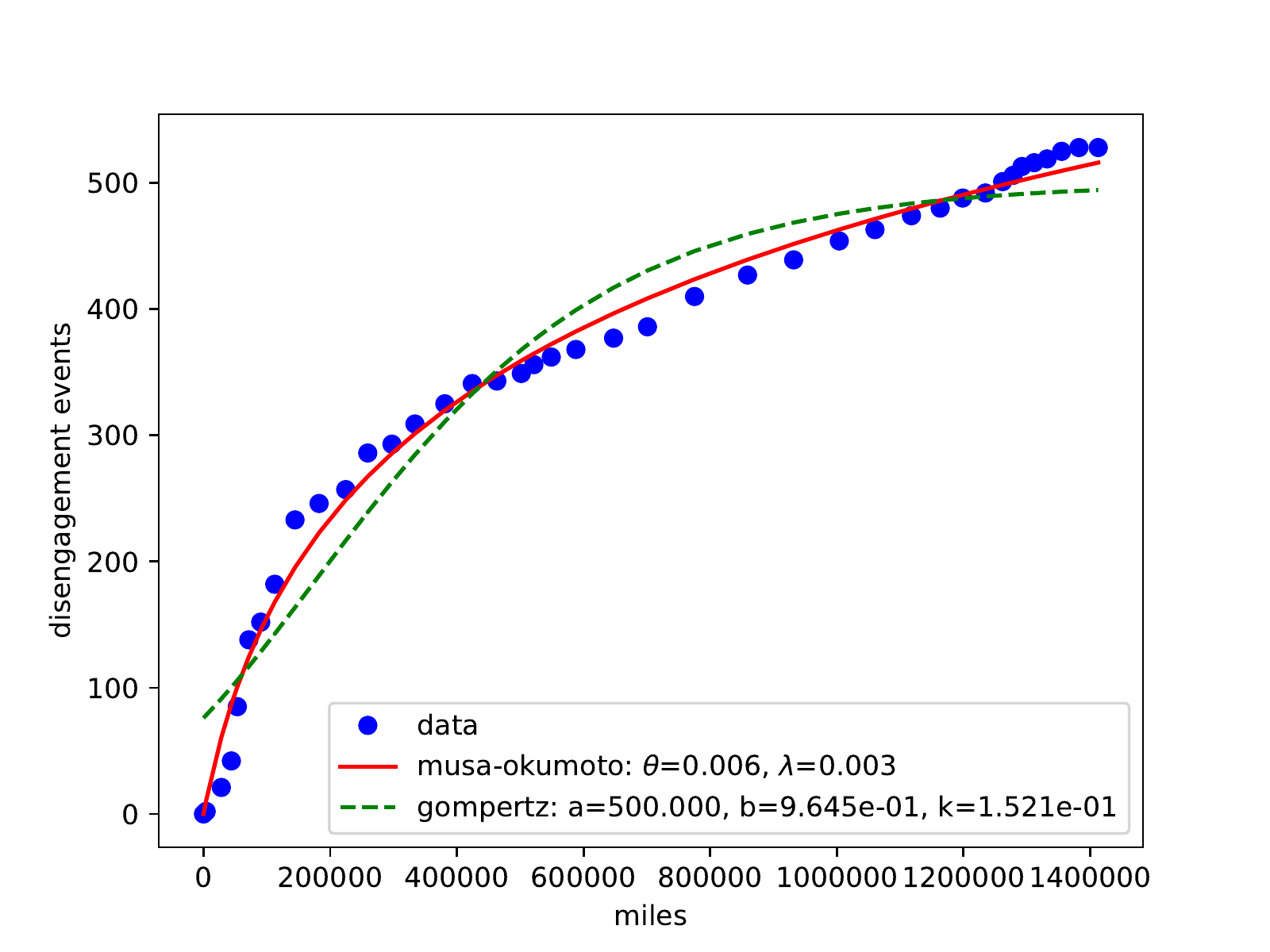}
\caption{{\bf Best-fit models for disengagement events for Waymo data set.}
}
\label{fig1}
\end{figure}

Fig~\ref{fig1} shows the observed cumulative number of disengagement events and kilometres
driven, and the least squares model fit for the Musa-Okumoto and Gompertz models for the Waymo data set.  Visually, the data set contains no indication of an S-shaped curve.  Visually, Musa-Okumoto model appears to be a reasonably good fit for the data; the Gompertz model less so.
As previously discussed, the use of nonlinear least squares estimation does not permit the calculation of confidence intervals for the model parameters, so there are no error bars
on the plot.

\begin{figure}[!h]
\insertpic{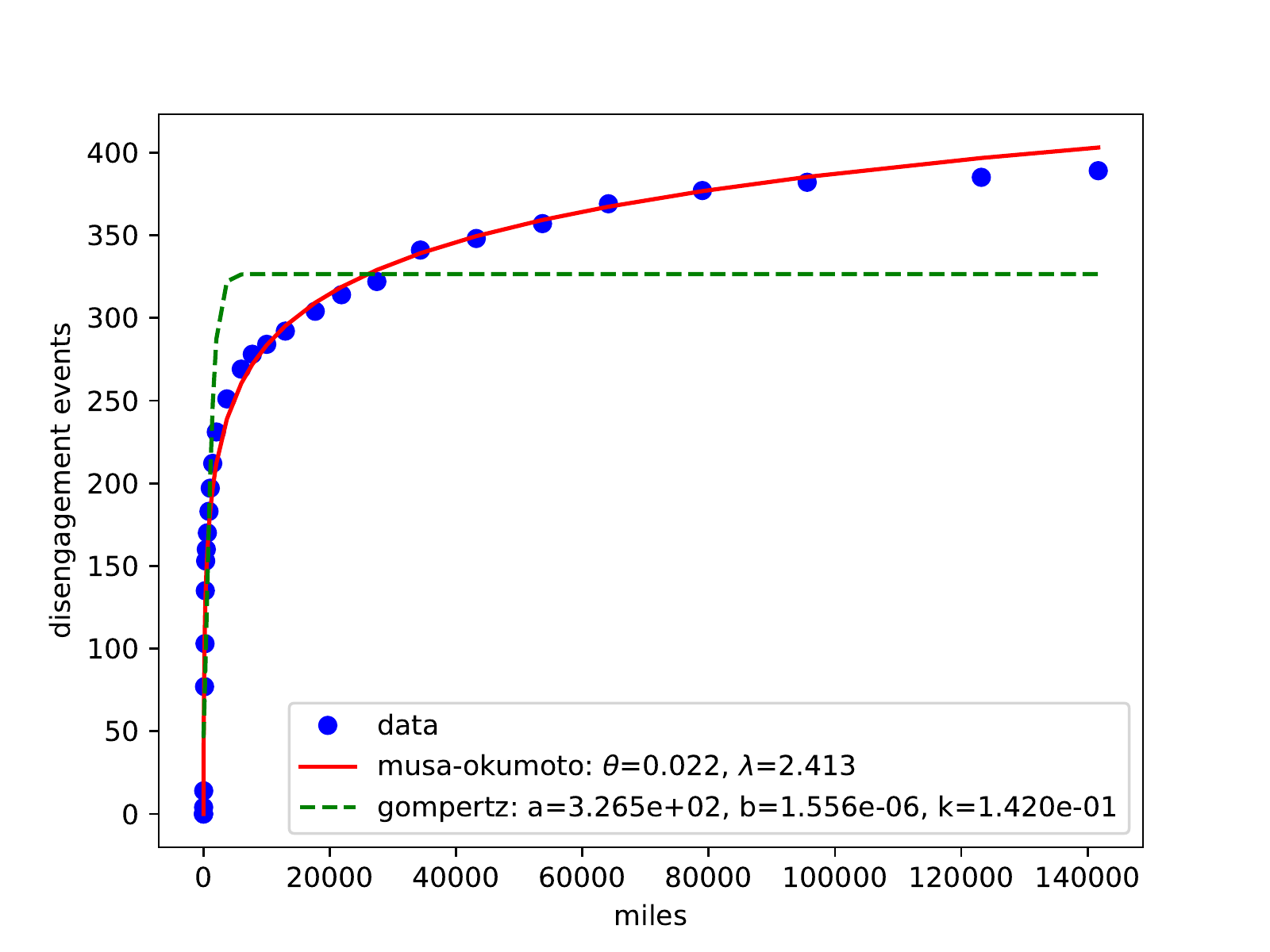}
\caption{{\bf Best-fit models for disengagement events for Cruise data set.}}
\label{fig2}
\end{figure}

Fig~\ref{fig2} shows the observed disengagement events and best fit models for the Cruise
data set. In this case, the Musa-Okumoto model is a far better visual fit than the Gompertz
model.  This may be due, in part, to the overrepresentation of points where the accumulated
distance is small in the model; the least squares estimator weights errors at all provided
data points equally, even if they are not equisdistributed on the x-axis.  Regardless, the
visual similarity of the best-fit Musa-Okumoto model to the observed data is striking.

Table \ref{table1} shows the standard error of the regression for each model and data set.  As
would be expected from the graphs, the standard error of estimate for the Musa-Okumoto
model is lower than for the Gompertz model, for both data sets.

\begin{figure}[!h]
\insertpic{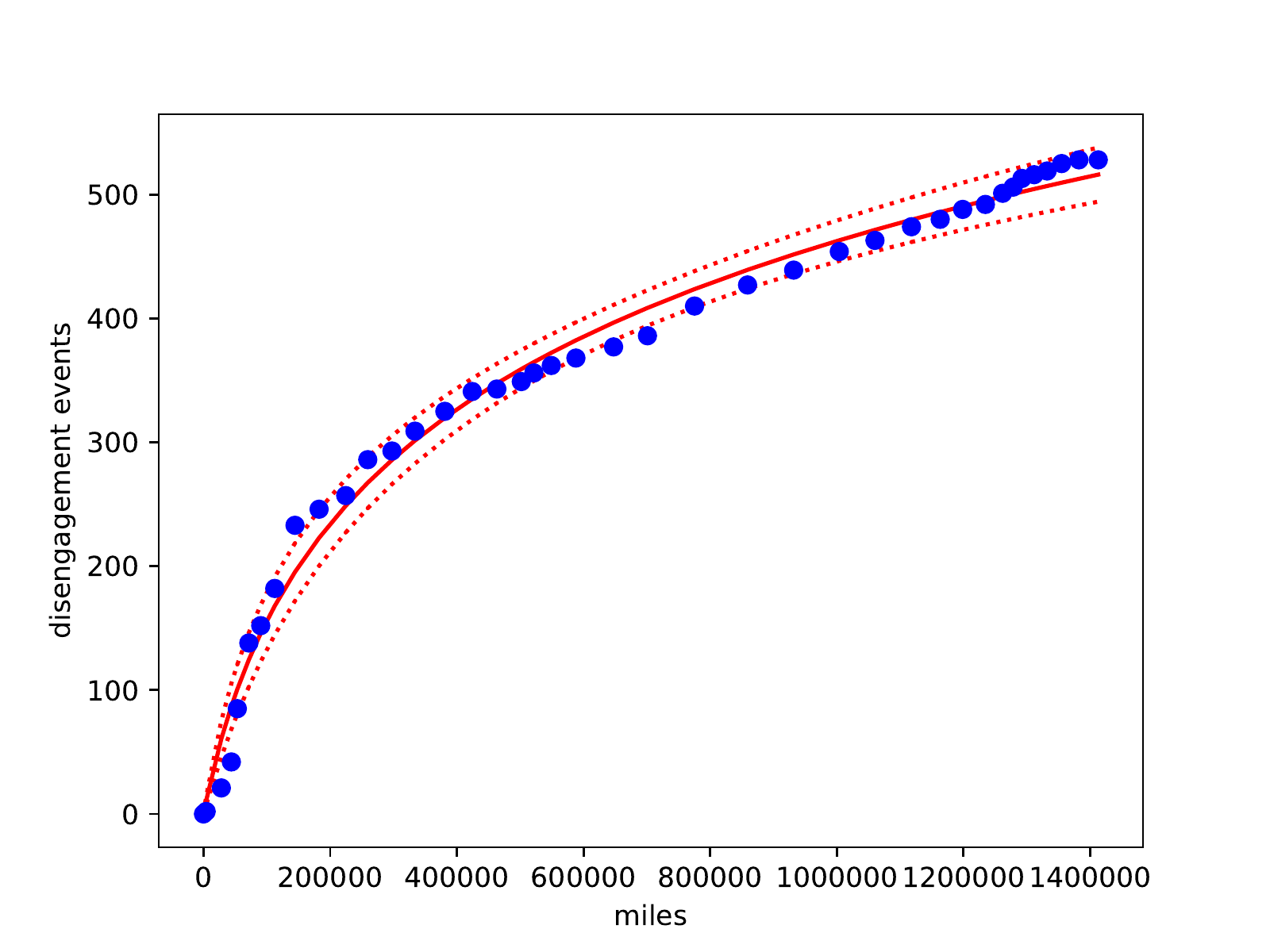}
\caption{{\bf Musa-Okumoto model fit, with confidence intervals for disengagement events for Waymo data set.}
}
\label{figgoogleconf}
\end{figure}

\begin{figure}[!h]
\insertpic{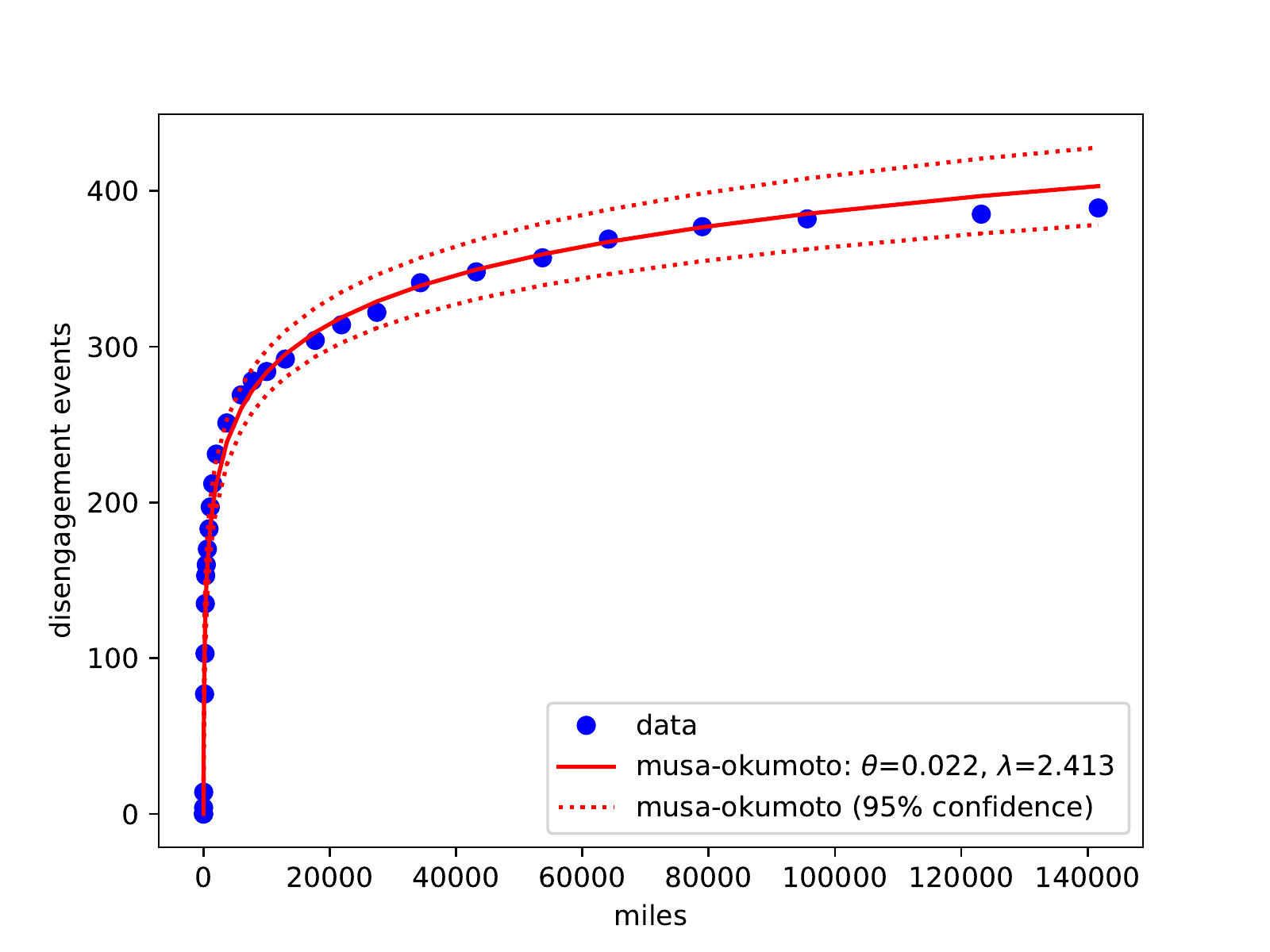}
\caption{{\bf Musa-Okumoto model fit, with confidence intervals for disengagement events for GM Cruise data set.}
}
\label{figcruiseconf}
\end{figure}

\begin{table}[!ht]
\centering
\caption{
{\bf Standard error of estimate for SRGMs }}
\begin{tabular}{|l||r|r|}
\hline
& \multicolumn{2}{|c|}{\bf Data set} \\
{\bf Model} & {\bf Waymo} & {\bf Cruise}\\ \hline \hline
Musa-Okumoto & 2.501 & 2.468 \\ \hline
Gompertz & 5.699 & 7.553  \\ \hline
\end{tabular}
\label{table1}

\end{table}
Figures~\ref{figgoogleconf} and~\ref{figcruiseconf} show 95\% confidence intervals for the
Musa-Okumoto model for the Waymo and Cruise datasets respectively.

\subsection{Experiment 2: prediction}

Attempts to find model parameters for a model of first two-thirds of the data points using the Gompertz model proved fruitless.  SciPy's curve fitting algorithm was unable to converge to a plausible solution for either data set.  Therefore, we present results for the Musa-Okumoto model only.

\begin{figure}[!h]
\insertpic{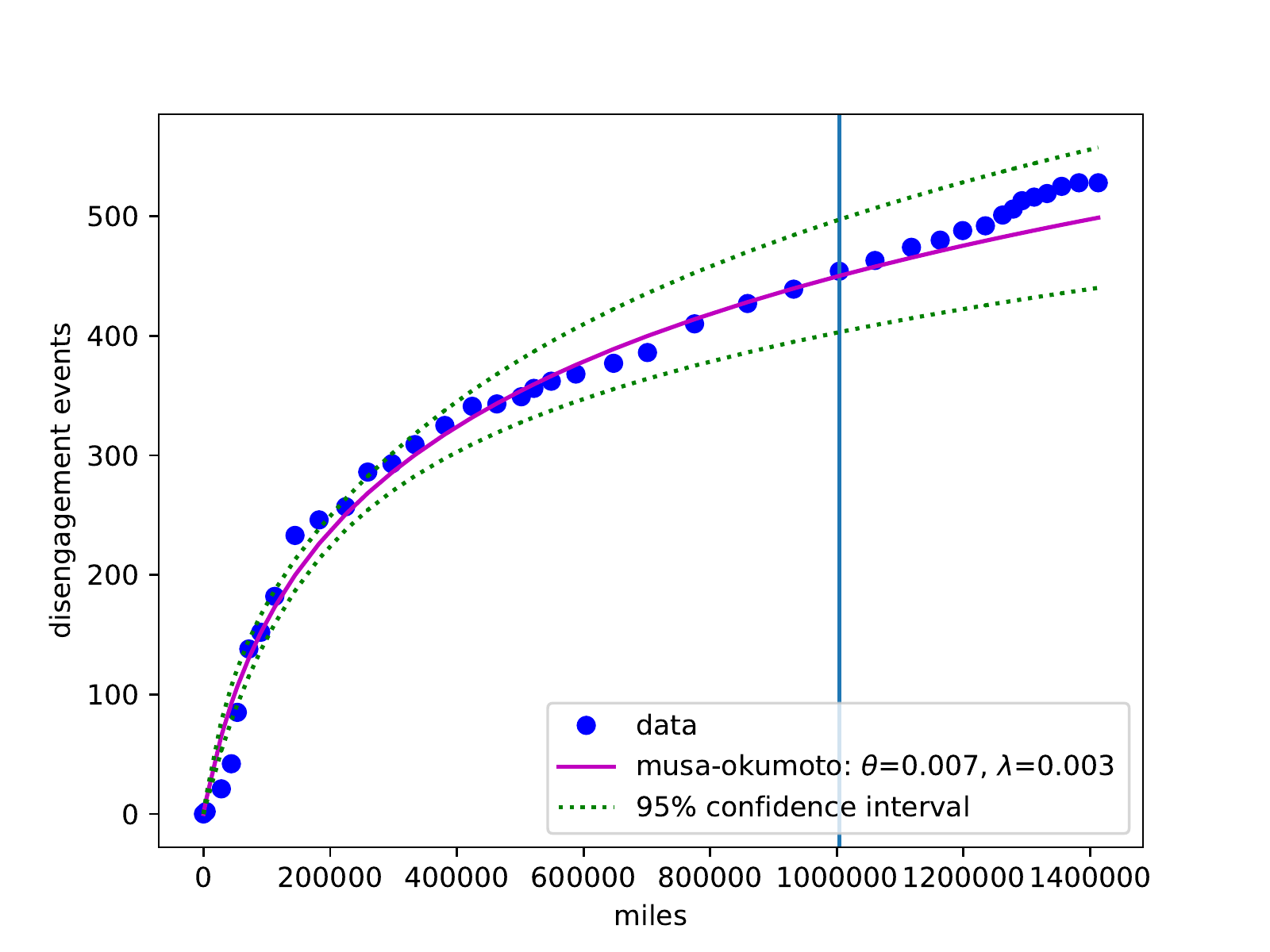}
\caption{{\bf Predictive model for Waymo data set}
Cumulative disengagement events for the Waymo data set compared to a best-fit model
computed from the first two-thirds of the monthly data points.}
\label{fig3}
\end{figure}

\begin{figure}[!h]
\insertpic{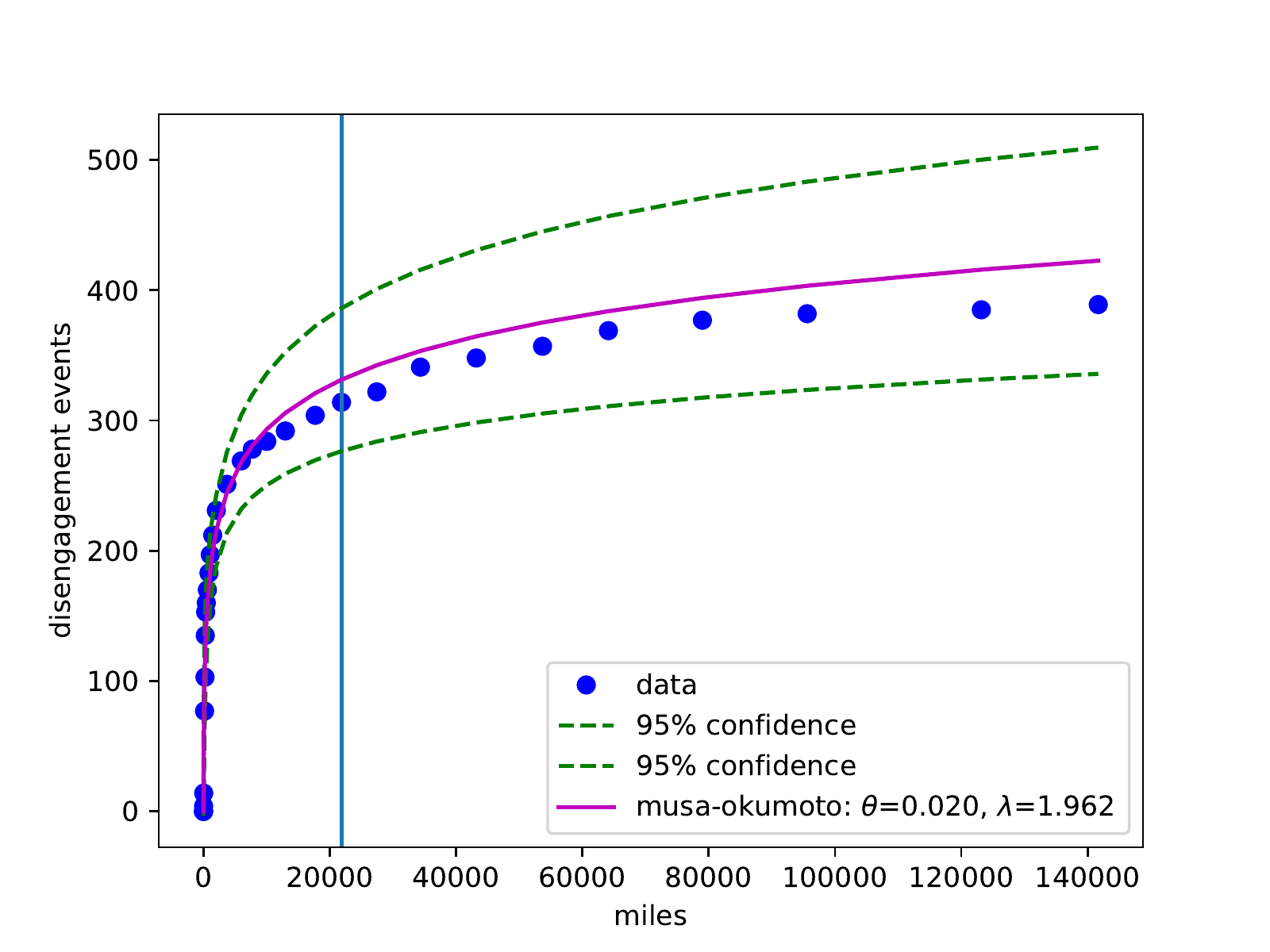}
\caption{{\bf Predictive model for Cruise data set}
Cumulative disengagement events for the Cruise data set compared to a best-fit model
computed from the first two-thirds of the monthly data points.}
\label{fig4}
\end{figure}

Figure \ref{fig3} shows a plot of actual disengagement events for Waymo, compared to a Musa-Okumoto model calculated from the first two-thirds of the monthly data points. Figure \ref{fig4} shows a similar plot for the Cruise.  Points left of the vertical line were in the first two-thirds of the monthly data points and were used in the model parameter estimation.  Note that the proportion of the testing kilometres included in each model differs substantially, as it appears Waymo reduced the number of testing kilometres driven in California in the latter
part of the data series, where Cruise increased their own testing.  Table \ref{table2} shows the standard error
of estimate for the two models.

\begin{table}[!ht]
\centering
\caption{
{\bf Standard error of estimate for SRGMs computed from first two-thirds of monthly data points }}
\begin{tabular}{|l||r|} \hline
{\bf Dataset} & Standard error of estimate \\ \hline
Waymo & 2.873  \\
Cruise & 2.771  \\ \hline
\end{tabular}
\label{table2}
%

\end{table}

In both cases, it appears that the actual data is broadly consistent
with the model predictions.  The model predicts fewer disengagement events than were present in the real data in the Waymo data set, but more events than actually occurred in the Cruise data set. It is notable that in both cases, but particularly for the Cruise dataset, the confidence intervals are relatively wide compared to the fits with the full datasets.

\section{Threats to validity}

There are a number of threats to the validity of this study.

The raw data presented in the paper was collected on behalf of subsidiaries
of large, publicly listed companies, who face legal sanction
if inaccurate or incomplete data was provided.  The calculations performed in this paper were conducted using the SciPy library, a well-tested, popular library for performing statistical modelling.  The (trivial) Python code used for the calculations, and the raw data, is available from (source).

However, there are a number of potential confounding factors that the data set provides
insufficient information to evaluate.  The models assumed that testing and debugging
practices did not change, and that the number of kilometres accumulated in California
testing linearly correlated with the total testing and debugging effort to date.  This is known not to be the full picture for Waymo.  Waymo conducts testing in a private testing ground~\cite{debord_waymo_2018},  performs
simulation experiments of variations of real-world events using \emph{fuzz testing}~\cite{debord_waymo_2018-1}, and was in the process of shifting much of its onroad testing to Arizona towards the end of the reporting period.  Unlike in California, in Arizona, companies are not required to disclose disengagement events, or accidents.   Fixes for faults found and located through
these efforts are presumably incorporated into the cars tested in California, but there
is little or no public information as to the extent and timelines of this private testing.  While less is known about the testing program of GM Cruise, they are highly
likely to also be conducting simulation testing, and testing on private roads.

It is also unknown whether the purpose and scope of on-road testing changed significantly in the periods covered by the data sets. It seems likely that the range of driving conditions
in which AV manufacturers would test their vehicles would be relatively narrow initially, and broaden considerably over time.  This broadened scope would presumably
include higher-risk driving conditions such as unfavourable weather (AV sensors are
known to be affected by snow~\cite{davies_clever_2016})), increased vehicular and pedestrian traffic, and poorer road conditions.  If problems
are detected in particular conditions, it is plausible that for a subsequent period testing will
focus on confirming the effectiveness of those problems, which may affect the disengagement rate for
that period.

The robustness of the estimates is somewhat constrained by the inability to use the standard
technique for parameter estimation in SRGMs - maximum likelihood estimation.  ML models do no assume that deviations from the model are normally distributed, and the use of non-standard
(in this domain) estimation techniques increased the risks of mistakes.

Obviously, with a data set of only two autonomous vehicle projects, it is unknown whether other
projects will follow a similar reliability growth curve.  Nor do the existing  data sets show a complete development life cycle, as neither vehicle had reached a point where the manufacturer
had sought approval for full autonomous operation without a safety driver.\footnote{According to media
reports, Waymo plans to launch a fully autonomous taxi services in Phoenix, Arizona, some time in late 2018~\cite{lee_fully_2018}.}  Therefore, it is not clear from the available data whether these SRGMs can provide accurate predictions all the way to the very high reliability levels required for safe commercial operation.

\section{Discussion}

These results clearly indicate that reported disengagement events in two large AV test programs
can be fitted reasonably accurately to standard software reliability growth models.  While not as accurate, the ability of the models to predict failure rates is also surprisingly good given all the uncertainties and confounding factors mentioned above.  The results also
demonstrate that confidence intervals for models can be calculated, allowing upper bounds
on disengagement rates, with nominated levels of confidence, can be calculated, on an
evolving system.  It is even possible to estimate how much further testing is required before
the upper bound disengagement rates will be below a desired threshold.

\subsection{Viability of reliability growth modeling in AVs}

The application of the Musa-Okumoto SRGM to the AV disengagement data suggests that, contrary
to the assumptions of Kalra and Paddock~\cite{kalra_driving_2016}, it may be possible to calculate mathematically sound estimates of the current and future failure rates of AVs based on past behaviour of earlier versions of the system.

Furthermore,  rather than trying to model comparatively rare events -- fatalities or accidents --
using disengagement events is a more feasible way to assess the the progress of an AV project. It should be noted that using self-reported disengagement events as an mandated threshold criterion for regulatory approval would be risky, as AV manufacturers might be tempted to modify their software and instruct their drivers to not disengage in marginal situations.

\subsection{Models}

For these two programs, the Musa-Okumoto model was a far better fit to the complete data sets, and was the only model that (at least with the SciPy's curve fitting algorithm) that resulted
in a plausible fit when provided with an incomplete data set.  The key assumption of an S-shaped model is that testers
initial efforts will be less effective at revealing faults than later efforts as they gain experience in testing the system.  The nature of road-based testing, and the nature of the faults reported is not consistent with this assumption, and the data appears to bear this out.

While the simple Musa-Okumoto model was a good fit for the data available, it is unlikely to be
the last word in modeling autonomous vehicle failures.  Software reliability growth models model
testing as a simple numeric quantity, but it is an open question as to whether using real-world testing kilometres travelled is appropriate here given the multifaceted nature of AV testing programs.  According to Madrigal~\cite{madrigal_inside_2017}, many of the
improvements in Waymo's AV program come through their use of simulation, rather than directly from
on-road testing.  However, the ``interesting'' scenarios they model are often ones identified in
disengagement scenarios.  As such, it is possible that simulation, powerful though it is, is ultimately an ``amplifier'' for real-world testing.  That is, it allows the exploration of a huge
number of variations of failure scenarios, and gives higher confidence that those failure
scenarios will not occur again when similar conditions are encountered in the real world.  If so,
it may well be that real-world testing kilometres travelled \emph{are} a reasonable proxy for testing effort, but it remains an open question that can only be clarified with more data.

\subsection{Data reporting procedure limitations}

As noted in threats to validity, the need to use nonlinear least squares estimation rather than maximum likelihood modeling has increased the risk of invalid assumptions in the models.  However, modifications to the data reporting procedure
would make it straightforward to compute maximum likelihood models for disengagement event rates. For instance, instead of reporting the number of failures per calendar month and the total distance driven, companies could report the cumulative distance driven by the test fleet at the time of each disengagement event.

\subsection{Related works}

In this context, we should also consider the work of Huang \etal~\cite{huang_accelerated_2017}, who observe
that testing driverless cars by reflecting real-world driving conditions is not a
particularly efficient way to improve or measure reliability.  Their approach
is to consider particular driving tasks (in their example, freeway lane changes), build a statistical model of the key parameters governing
the variance in lane changes, skew testing (both physical and simulation) heavily to those areas of the parameter space where failures are likely to occur, and then use the statistical model in reverse to estimate real-world failure rates.  SRGMs could,
in theory, be used as part of such an approach, as a way to track the decrease in failure rates in specific tasks. It may also be considered desirable to conduct a
parallel on-road testing replicating normal driving to validate any Huang-style
statistical model - and using an SRGM on this data would still be useful as a way
to provide ongoing estimates of vehicle failure rates, to compare with predictions.

Favaro et al~\cite{favaro_examining_2017} examined accident data for the Waymo program through to 2017, and found a simple linear relationship between kilometres travelled and cumulative accidents.  They therefore concluded,  While accidents are an important metric, it does not follow that there have been no improvements in
the function of AV systems.  Their analysis does does not take into account that manual intervention by the human safety drivers is likely to have prevented a significant number of accidents, and those disengagements, as shown in this work,
have become much rarer over time.  Furthermore, it does not take into account the contribution to accidents of human drivers that an AV could not reasonably be expected to avoid.  Favaro \etal~ also examined the circumstances of disengagement
events in some detail.

The present paper, along with Favaro and others who have examined the California
DMV data, shows the value of open data sets for enabling research into topics of
public importance.  However, it also demonstrates the need for expert advice on
exactly what data is collected and reported.  If the advice of statisticians had been
sought, it may well be that cumulative distances driven at the time of disengagement
events would have been included in the data set, making maximum
likelihood estimates straightforward to calculate.

\section{Conclusion}

The key contributions of this paper are as follows:
\begin{enumerate}
 \item Software reliability growth models (SRGMs) closely fit trends in disengagement events in the two most extensive public road AV test programs in the California DMV public data set, and can therefore be used to rigorously estimate the current disengagement rate.
 \item Software reliability growth models are reasonable predictors of trends in disengagement events in these two programs.  This could allow an AV manufacturer or other interested party
 to predict the expected future disengagement rate, given a specified amount of continued testing and development.
 \item Confidence intervals for model parameters and the model values can be calculated,
 enabling the estimation of the present or future probability that disengagement rates are
 lower than a desired threshold.
 \item A representative concave SRGM, Musa-Okumoto, is a better fit
 to the data than a representative S-shaped SRGM, the Gompertz model.
 \item The mandated format of the California DMV data set does not permit the
 straightforward use of maximum likelihood model estimation, which is a more robust
 way of estimating the parameters of an SRGM.
\end{enumerate}

These contributions are of value not only to developers of AV systems, who may be
able to use SRGMs to evaluate the present state of an AV system, but also to estimate
the time and resources required to achieve safety benchmarks before those resources are invested.  They also have implications for regulators, as it offers the potential for
rigorous estimates of AV system safety without requiring infeasible amounts of testing.

There are many potential extensions to this preliminary work.  As more AV programs
mature, an obvious followup is to examine whether the disengagement events for those programs are also effectively modelled using SRGMs.  With more data, it should
also be possible to consider more rigorously whether the Musa-Okumoto model is really
the most applicable to AV road testing data, or if one or more of the many other concave SRGMs is a better fit.  It would also be beneficial to investigate integrating the use of
SRGMs into the weighted modelling/testing approach of Huang \etal~\cite{huang_accelerated_2017}.

As well as ML models, There are other candidate techniques for parameter estimation and calculation of confidence intervals, which may be applicable to this data.  Bootstrapping~\cite{lahiri_resampling_2003} can be used to calculate variance for certain
types of time-dependent data.  Bootstrapping, and other resampling techniques, may permit
more accurate confidence intervals to be calculated on the relatively small data sets
available here.

The relationship between disengagement events and accidents, the metric of ultimate concern, is also a topic worthy of much further work.  While the question of assigning responsibility in accidents is knotty, a working understanding must be
found if regulators are to fairly assess the safety of AV systems without unfairly penalizing them for actions of others that human drivers would not be held accountable
for.  This multifaceted problem will require an interdisciplinary solution, with the social sciences and humanities having at least as much to contribute as science and engineering.

While this paper has concentrated on land-based autonomous vehicles, other types of robotic vehicles face similar concerns.  There is currently considerable commercial interest in
developing uncrewed, fully autonomous drones for package delivery, and much larger autonomous aircraft for passenger transport~\cite{sorkin_larry_2018}.  In both cases, there is a
need for reliability estimation and prediction, and it is plausible that SRGMs will be useful
for this purpose.  Therefore, performing a similar analysis to this paper with a drone failure
dataset is an opportunity for future work with near-term real-world implications.

In the race to commercialize what is expected to be exceedingly
valuable technology, questionable safety practices have been revealed~\cite{merkel_preliminary_2018} in some AV development programs.  Simply trusting assertions that an AV is sufficiently safe is unlikely to, and should not, satisfy the public as to the readiness of an AV system for production use.
Whatever the ultimate statistical basis of demonstrating the reliability and
safety of autonomous vehicles, it is clear that an explicit, rigorous and statistically sound approach to safety assessment by regulators is required.
This paper is, hopefully, a small contribution towards that.

\appendices


\ifCLASSOPTIONcompsoc
  \section*{Acknowledgments}
\else
  \section*{Acknowledgment}
\fi

The author would like to thank John Grundy and Marcel Boehme for their helpful comments on
an earlier draft of this paper.

\ifCLASSOPTIONcaptionsoff
  \newpage
\fi



\bibliographystyle{IEEEtran}
\bibliography{mylibrary}

\begin{thebibliography}{10}
\providecommand{\url}[1]{#1}
\csname url@samestyle\endcsname
\providecommand{\newblock}{\relax}
\providecommand{\bibinfo}[2]{#2}
\providecommand{\BIBentrySTDinterwordspacing}{\spaceskip=0pt\relax}
\providecommand{\BIBentryALTinterwordstretchfactor}{4}
\providecommand{\BIBentryALTinterwordspacing}{\spaceskip=\fontdimen2\font plus
\BIBentryALTinterwordstretchfactor\fontdimen3\font minus
  \fontdimen4\font\relax}
\providecommand{\BIBforeignlanguage}[2]{{%
\expandafter\ifx\csname l@#1\endcsname\relax
\typeout{** WARNING: IEEEtran.bst: No hyphenation pattern has been}%
\typeout{** loaded for the language `#1'. Using the pattern for}%
\typeout{** the default language instead.}%
\else
\language=\csname l@#1\endcsname
\fi
#2}}
\providecommand{\BIBdecl}{\relax}
\BIBdecl

\bibitem{noauthor_automated_nodate-1}
\BIBentryALTinterwordspacing
``Automated vehicles in {Australia},'' National Transport Commission, Tech.
  Rep. [Online]. Available:
  \url{https://www.ntc.gov.au/roads/technology/automated-vehicles-in-australia/}
\BIBentrySTDinterwordspacing

\bibitem{liu_how_2018}
\BIBentryALTinterwordspacing
P.~Liu, R.~Yang, and Z.~Xu, ``\BIBforeignlanguage{en}{How {Safe} {Is} {Safe}
  {Enough} for {Self}-{Driving} {Vehicles}?: {How} {Safe} {Is} {Safe} {Enough}
  for {Self}-{Driving} {Vehicles}},'' \emph{\BIBforeignlanguage{en}{Risk
  Analysis}}, May 2018. [Online]. Available:
  \url{http://doi.wiley.com/10.1111/risa.13116}
\BIBentrySTDinterwordspacing

\bibitem{kalra_driving_2016}
\BIBentryALTinterwordspacing
N.~Kalra and S.~M. Paddock, ``\BIBforeignlanguage{en}{Driving to safety: {How}
  many miles of driving would it take to demonstrate autonomous vehicle
  reliability?}'' \emph{\BIBforeignlanguage{en}{Transportation Research Part A:
  Policy and Practice}}, vol.~94, pp. 182--193, Dec. 2016. [Online]. Available:
  \url{http://linkinghub.elsevier.com/retrieve/pii/S0965856416302129}
\BIBentrySTDinterwordspacing

\bibitem{harris_google_2017}
\BIBentryALTinterwordspacing
M.~Harris, ``\BIBforeignlanguage{en}{Google {Has} {Spent} {Over} \$1.1
  {Billion} on {Self}-{Driving} {Tech}},'' Sep. 2017. [Online]. Available:
  \url{https://spectrum.ieee.org/cars-that-think/transportation/self-driving/google-has-spent-over-11-billion-on-selfdriving-tech}
\BIBentrySTDinterwordspacing

\bibitem{wood_software_1996}
\BIBentryALTinterwordspacing
A.~Wood, ``\BIBforeignlanguage{en}{Software {Reliability} {Growth} {Models}},''
  Tandem, Technical {Report} 96.1, 1996, part number 130056. [Online].
  Available: \url{http://www.hpl.hp.com/techreports/tandem/TR-96.1.pdf}
\BIBentrySTDinterwordspacing

\bibitem{rechnitzer_george_effect_2000}
\BIBentryALTinterwordspacing
G.~Rechnitzer, N.~Haworth, and N.~Kowadlo, ``The {Effect} of {Vehicle}
  {Roadworthiness} on {Crash} {Incidence} and {Severity},'' Monash University,
  Monash University Accident Research Centre, Technical {Report} 164, 2000,
  sponsoring Organisation(s): The Victorian Automobile Chamber of Commerce.
  [Online]. Available:
  \url{https://www.monash.edu/__data/assets/pdf_file/0017/216710/muarc164.pdf}
\BIBentrySTDinterwordspacing

\bibitem{favaro_examining_2017}
\BIBentryALTinterwordspacing
F.~M. Favarò, N.~Nader, S.~O. Eurich, M.~Tripp, and N.~Varadaraju, ``Examining
  accident reports involving autonomous vehicles in {California},'' \emph{PLOS
  ONE}, vol.~12, no.~9, pp. 1--20, 2017. [Online]. Available:
  \url{https://doi.org/10.1371/journal.pone.0184952}
\BIBentrySTDinterwordspacing

\bibitem{noauthor_testing_nodate}
\BIBentryALTinterwordspacing
``Testing of {Autonomous} {Vehicles}.'' [Online]. Available:
  \url{https://www.dmv.ca.gov/portal/dmv/detail/vr/autonomous/testing}
\BIBentrySTDinterwordspacing

\bibitem{noauthor_taxonomy_2018}
\BIBentryALTinterwordspacing
``Taxonomy and {Definitions} for {Terms} {Related} to {Driving} {Automation}
  {Systems} for {On}-{Road} {Motor} {Vehicles},'' Jun. 2018. [Online].
  Available: \url{https://www.sae.org/standards/content/j3016_201806/}
\BIBentrySTDinterwordspacing

\bibitem{noauthor_ieee_2017}
``{IEEE} {Recommended} {Practice} on {Software} {Reliability},'' \emph{IEEE Std
  1633-2016 (Revision of IEEE Std 1633-2008)}, pp. 1--261, Jan. 2017.

\bibitem{musa_john_d._software_1987}
J.~D. Musa, A.~Iannino, and K.~Okumoto, \emph{Software {Reliability}:
  {Measurement}, {Prediction}, {Application}}.\hskip 1em plus 0.5em minus
  0.4em\relax McGraw-Hill, 1987.

\bibitem{ullah_comparative_2012}
\BIBentryALTinterwordspacing
N.~Ullah, M.~Morisio, and A.~Vetro, ``A {Comparative} {Analysis} of {Software}
  {Reliability} {Growth} {Models} using {Defects} {Data} of {Closed} and {Open}
  {Source} {Software},'' in \emph{2012 35th {Annual} {IEEE} {Software}
  {Engineering} {Workshop}}.\hskip 1em plus 0.5em minus 0.4em\relax IEEE, Oct.
  2012, pp. 187--192. [Online]. Available:
  \url{http://ieeexplore.ieee.org/document/6479816/}
\BIBentrySTDinterwordspacing

\bibitem{musa_logarithmic_1984}
J.~D. Musa and K.~Okumoto, ``A logarithmic {Poisson} execution time model for
  software reliability measurement,'' in \emph{Proceedings of the 7th
  international conference on {Software} engineering}.\hskip 1em plus 0.5em
  minus 0.4em\relax IEEE Press, 1984, pp. 230--238.

\bibitem{huang_unified_2003}
C.-Y. Huang, M.~R. Lyu, and S.-Y. Kuo, ``A unified scheme of some
  {Nonhomogenous} {Poisson} process models for software reliability
  estimation,'' \emph{IEEE Transactions on Software Engineering}, vol.~29,
  no.~3, pp. 261--269, Mar. 2003.

\bibitem{noauthor_waymo_nodate}
\BIBentryALTinterwordspacing
``\BIBforeignlanguage{en}{Waymo}.'' [Online]. Available:
  \url{https://waymo.com/}
\BIBentrySTDinterwordspacing

\bibitem{noauthor_cruise_nodate}
\BIBentryALTinterwordspacing
``Cruise {Automation}.'' [Online]. Available: \url{https://www.getcruise.com/}
\BIBentrySTDinterwordspacing

\bibitem{pawitan_yudi_all_2001}
Y.~Pawitan, \emph{In all likelihood : statistical modelling and inference using
  likelihood}, 1st~ed.\hskip 1em plus 0.5em minus 0.4em\relax Oxford: Clarendon
  Press, 2001.

\bibitem{noauthor_scipy_2018}
\BIBentryALTinterwordspacing
``{SciPy},'' May 2018. [Online]. Available:
  \url{https://docs.scipy.org/doc/scipy/reference/index.html}
\BIBentrySTDinterwordspacing

\bibitem{noauthor_how_nodate}
\BIBentryALTinterwordspacing
``How does {Prism} compute confidence and prediction bands for nonlinear
  regression or linear regression? - {FAQ} 1099 - {GraphPad}.'' [Online].
  Available: \url{https://www.graphpad.com/support/faqid/1099/}
\BIBentrySTDinterwordspacing

\bibitem{spiess_evaluation_2010}
\BIBentryALTinterwordspacing
A.-N. Spiess and N.~Neumeyer, ``An evaluation of {R}2 as an inadequate measure
  for nonlinear models in pharmacological and biochemical research: a {Monte}
  {Carlo} approach,'' \emph{BMC Pharmacology}, vol.~10, p.~6, Jun. 2010.
  [Online]. Available:
  \url{https://www.ncbi.nlm.nih.gov/pmc/articles/PMC2892436/}
\BIBentrySTDinterwordspacing

\bibitem{salkind_standard_2010}
\BIBentryALTinterwordspacing
``Standard {Error} of {Estimate},'' in \emph{Encyclopedia of {Research}
  {Design}}.\hskip 1em plus 0.5em minus 0.4em\relax 2455 Teller Road, Thousand
  Oaks California 91320 United States: SAGE Publications, Inc., 2010.
  [Online]. Available:
  \url{http://methods.sagepub.com/reference/encyc-of-research-design/n435.xml}
\BIBentrySTDinterwordspacing

\bibitem{debord_waymo_2018}
\BIBentryALTinterwordspacing
M.~Debord, ``A {Waymo} employee reveals how the company is putting their
  self-driving cars to the test,'' Aug. 2018. [Online]. Available:
  \url{https://www.businessinsider.com.au/waymo-alphabet-employee-reveals-how-self-driving-cars-are-tested-2018-8?r=US&IR=T}
\BIBentrySTDinterwordspacing

\bibitem{debord_waymo_2018-1}
\BIBentryALTinterwordspacing
------, ``A {Waymo} engineer told us why a virtual-world simulation is crucial
  to the future of self-driving cars,'' Aug. 2018. [Online]. Available:
  \url{https://www.businessinsider.com.au/waymo-engineer-explains-why-testing-self-driving-cars-virtually-is-critical-2018-8?r=US&IR=T}
\BIBentrySTDinterwordspacing

\bibitem{davies_clever_2016}
\BIBentryALTinterwordspacing
A.~Davies, ``The {Clever} {Way} {Ford}'s {Self}-{Driving} {Cars} {Navigate} in
  {Snow},'' Nov. 2016. [Online]. Available:
  \url{https://www.wired.com/2016/01/the-clever-way-fords-self-driving-cars-navigate-in-snow/}
\BIBentrySTDinterwordspacing

\bibitem{lee_fully_2018}
\BIBentryALTinterwordspacing
T.~Lee, ``Fully driverless {Waymo} taxis are due out this year, alarming
  critics,'' Oct. 2018. [Online]. Available:
  \url{https://arstechnica.com/cars/2018/10/waymo-wont-have-to-prove-its-driverless-taxis-are-safe-before-2018-launch/}
\BIBentrySTDinterwordspacing

\bibitem{madrigal_inside_2017}
\BIBentryALTinterwordspacing
A.~Madrigal, ``Inside {Waymo}'s {Secret} {World} for {Training}
  {Self}-{Driving} {Cars},'' Aug. 2017. [Online]. Available:
  \url{https://www.theatlantic.com/technology/archive/2017/08/inside-waymos-secret-testing-and-simulation-facilities/537648/}
\BIBentrySTDinterwordspacing

\bibitem{huang_accelerated_2017}
\BIBentryALTinterwordspacing
Z.~Huang, D.~Zhao, H.~Lam, and D.~J. LeBlanc,
  ``\BIBforeignlanguage{en}{Accelerated {Evaluation} of {Automated} {Vehicles}
  {Using} {Piecewise} {Mixture} {Models}},''
  \emph{\BIBforeignlanguage{en}{arXiv:1701.08915 [cs]}}, Jan. 2017, arXiv:
  1701.08915. [Online]. Available: \url{http://arxiv.org/abs/1701.08915}
\BIBentrySTDinterwordspacing

\bibitem{lahiri_resampling_2003}
S.~N. Lahiri, \emph{Resampling {Methods} for {Dependent} {Data}}, ser. Springer
  {Series} in {Statistics}.\hskip 1em plus 0.5em minus 0.4em\relax
  Springer-Verlag, 2003.

\bibitem{sorkin_larry_2018}
\BIBentryALTinterwordspacing
A.~R. Sorkin, ``\BIBforeignlanguage{en-US}{Larry {Page}’s {Flying} {Taxis},
  {Now} {Exiting} {Stealth} {Mode}},'' \emph{\BIBforeignlanguage{en-US}{The New
  York Times}}, Jul. 2018. [Online]. Available:
  \url{https://www.nytimes.com/2018/03/12/business/dealbook/flying-taxis-larry-page.html}
\BIBentrySTDinterwordspacing

\bibitem{merkel_preliminary_2018}
\BIBentryALTinterwordspacing
R.~Merkel, ``\BIBforeignlanguage{en}{Preliminary report on {Uber}'s driverless
  car fatality shows the need for tougher regulatory controls},'' May 2018.
  [Online]. Available:
  \url{http://theconversation.com/preliminary-report-on-ubers-driverless-car-fatality-shows-the-need-for-tougher-regulatory-controls-97253}
\BIBentrySTDinterwordspacing

\end{thebibliography}

%

\begin{IEEEbiography}{Robert Merkel}
Robert is a Lecturer in software engineering at the Faculty of Information Technology at Monash University, Australia.
His primary research interests are in software quality and reliability, and
encompasses both the technical and human aspects of these topics.  He completed
his PhD at Swinburne University of Technology.
\end{IEEEbiography}







\end{document}